\pdfoutput=1
\documentclass[aps,prl,floatfix,superscriptaddress,reprint]{revtex4-2}
\usepackage{eurosym}
\usepackage{amsbsy}
\usepackage{latexsym,epsfig,graphicx}
\usepackage{dcolumn}
\usepackage{subfigure}
\usepackage{comment}
\usepackage{color}
\usepackage{bm}
\usepackage{mathrsfs}
\usepackage{amssymb}
\usepackage{amsfonts}
\usepackage{amsmath}
\usepackage{xspace}
\usepackage{epstopdf}
\usepackage{tabularx}
\usepackage{longtable}
\usepackage[colorlinks=true, letterpaper=true, pdfstartview=FitV, linkcolor=red, citecolor=blue, urlcolor=blue]{hyperref}
\usepackage[normalem]{ulem}
\usepackage[version=4]{mhchem}

\begin{document}
\title{Knot Topology of Exceptional Point and Non-Hermitian No-Go Theorem}
\author{Haiping Hu}
\email{hhu@iphy.ac.cn}
\affiliation{Beijing National Laboratory for Condensed Matter Physics, Institute of Physics, Chinese Academy of Sciences, Beijing 100190, China}
\author{Shikang Sun}
\affiliation{Beijing National Laboratory for Condensed Matter Physics, Institute of Physics, Chinese Academy of Sciences, Beijing 100190, China}
\affiliation{School of Physical Sciences, University of Chinese Academy of Sciences, Beijing 100049, China}
\author{Shu Chen}
\affiliation{Beijing National Laboratory for Condensed Matter Physics, Institute of Physics, Chinese Academy of Sciences, Beijing 100190, China}
\affiliation{School of Physical Sciences, University of Chinese Academy of Sciences, Beijing 100049, China}
\affiliation{Yangtze River Delta Physics Research Center, Liyang, Jiangsu 213300, China}

\begin{abstract}
Exceptional points (EPs) are peculiar band singularities and play a vital role in a rich array of unusual optical phenomena and non-Hermitian band theory. In this paper, we provide a topological classification of isolated EPs based on homotopy theory. In particular, the classification indicates that an $n$-th order EP in two dimensions is fully characterized by the braid group B$_n$, with its eigenenergies tied up into a geometric knot along a closed path enclosing the EP. The quantized discriminant invariant of the EP is the \textit{writhe} of the knot. The knot \textit{crossing number} gives the number of bulk Fermi arcs emanating from each EP. Furthermore, we put forward a non-Hermitian \textit{no-go} theorem, which governs the possible configurations of EPs and their splitting rules on a two-dimensional lattice and goes beyond the previous fermion doubling theorem. We present a simple algorithm generating the non-Hermitian Hamiltonian with a prescribed knot. Our framework constitutes a systematic topological classification of the EPs and paves the way towards exploring the intriguing phenomena related to the enigmatic non-Hermitian band degeneracy.
\end{abstract}
\maketitle

{\color{blue}\textit{Introduction.}} Non-Hermitian (NH) systems are ubiquitous in physics \cite{coll1,coll2,coll3,coll4,coll5,coll6,colladd1,colladd2,colladd3}, as epitomized by various photonic platforms with gain and loss \cite{op1,op2,op3,op4,op5,op6,xuepengep,xuepengep2,op7,op8,op9,op10,op11,op12,op13,coll4,nhwsmhhp,James,eti}. One of the most remarkable features of NH systems is that they exhibit level degeneracy of their complex eigenenergies, called exceptional points (EPs) \cite{ep,ep2}.  At an EP, two or more eigenvalues, and their corresponding eigenvectors, simultaneously coalesce (i.e., the Hamiltonian is defective), giving rise to phenomena without any analogs in the Hermitian realm. The intriguing properties of EPs have been widely exploited such as in unconventional transmission or reflection \cite{linz}, enhancing sensing \cite{sense1,sense2,sense3}, single-mode lasing \cite{lasing1,lasing2} and non-reciprocal phase transitions \cite{nrpt}.

EPs can be categorized into different types \cite{epclass}. An EP is of $n$-th order or $n$-fold (EP$_n$) if $n$ eigenstates simultaneously coalescence, i.e., the Hamiltonian is diagonalized into an $n\times n$ Jordan normal form. Similar to the well-known Dirac point or Weyl point in Hermitian systems, EP can be characterized by assigning an integer invariant like the vorticity of eigenvalues \cite{fuliang} or discriminant number \cite{yangzhesen}. Despite the explosive theoretical and experimental researches of exceptional degeneracies during the past few years, a comprehensive understanding of their exotic features and topological classification has been achieved yet. The characterization using discriminant number is far from satisfactory and incomplete. First, EPs of different types (especially of higher-order) that cannot smoothly evolve from one to another may have the same integer invariant introduced before. Second, NH Hamiltonians generally have complex eigenenergies compared to Hermitian systems, bringing critical differences in their topological characterizations \cite{class1,class2,knotPRL,knotexp,wangzhongnhse}. Near an EP, the energy levels may braid and tangle together and a full description of EPs requires the information of their nearby braiding patterns and branch cuts, hence they cannot be captured solely by simple integer invariants. Moreover, the richness of the band braiding near an EP may induce novel types of EPs previously unidentified.

In this paper, we formulate a homotopic classification of EPs in arbitrary dimensional NH systems (or generic parameter space), which enables the calculation of topological invariants using the machinery well-developed in algebraic topology. Particularly in 2D, our main findings are: \textit{i)} An isolated $n$-th order EP is fully characterized by braid group B$_n$, which reduces to integer group $\mathbb{Z}$ for EP$_2$. Our classification reveals the existence of infinitely many kinds of EPs which one-to-one correspond to the geometry knots. \textit{ii)} We demonstrate that the quantized vorticity (or discriminant number) and the number of bulk Fermi arcs emanating from the EP is the \textit{writhe} and \textit{crossing number} of the knot, respectively. \textit{iii)} Based on the classification, we put forward a NH no-go theorem that goes beyond the previous fermion doubling theorem and dictates the possible configurations of EPs and their splitting rules on a 2D lattice. \textit{iv)} To facilitate experimental realizations, we show how to generate a NH Hamiltonian hosting an EP of a prescribed knot pattern. Our framework bridging the NH physics, algebraic topology and knot theory, manifests the beauty and diversity of EPs and opens a broad avenue for investigating the exotic features of NH band degeneracies.
\begin{figure*}[t!]
\centering
\includegraphics[width=0.98\textwidth]{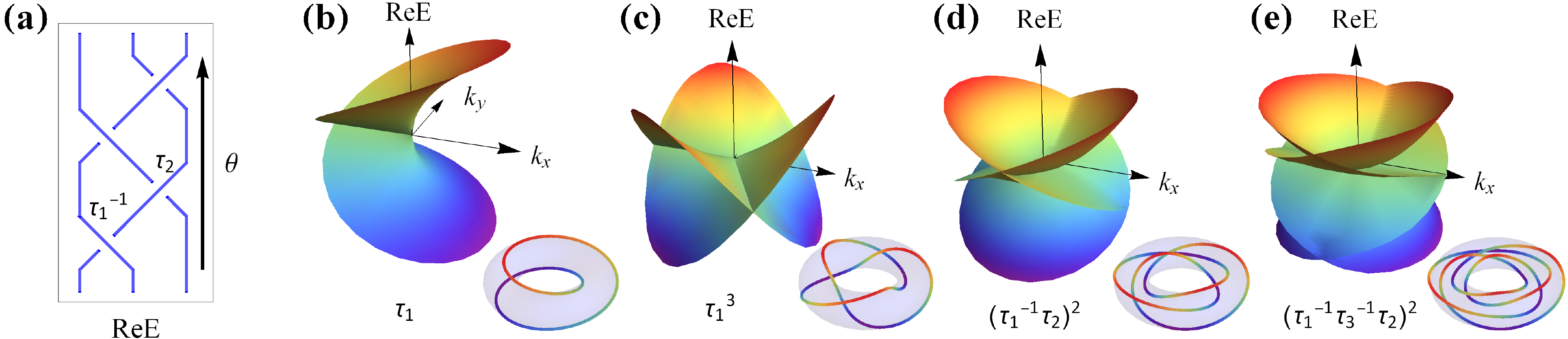}
\caption{Exceptional points (EPs) with energy-level braiding and knot topology. (a) Braid diagram marked by Artin's braid word notation: $\tau_i$ ($\tau_i^{-1}$) represents the $i$-th strand crosses over/under the $i+1$-th strand when traveling upwards. Closure of the braid by identifying the up and bottom ends forms a knot (in this case, it is the figure-8 knot). (b)-(e) Band structures (real part) and their associated knots along a closed path enclosing the EP (lower panels) for some representative EPs. (b) EP$_2$ of unknot, braid word: $\tau_1$, the characteristic polynomial (ChP) is $\lambda^2-z$, with $z=k_x+ik_y$; (c) EP$_2$ of trefoil knot, braid word: $\tau_1^3$, the ChP is $\lambda^2-z^3$; (d) EP$_3$ of figure-8 knot, braid word $(\tau_1^{-1}\tau_2)^2$. The ChP for figure-8 knot is given by $f_{F8}=1/64[64\lambda^3-12\lambda a^2[3(z\bar{z})^2+2z^2(z\bar{z})-2\bar{z}^2(z\bar{z})]-14a^3(z^2+\bar{z}^2)(z\bar{z})^2-a^3(z^4-\bar{z}^4)(z\bar{z})]$ with $a=1/4$; (e) EP$_4$ of $L6a1$ link, braid word: $(\tau_1^{-1}\tau_3^{-1}\tau_2)^2$.}\label{EPknot}
\end{figure*}

{\color{blue}\textit{Homotopy classification.}} Let us consider an $n$-band NH Hamiltonian $H(\bm k)=H(k_1,k_2,...,k_d)$ of $d$D system (or parameter space) and suppose an isolated $n$-fold EP located at $\bm k=0$. At the EP, the Hamiltonian can be brought to the Jordan normal form (For simplicity, the EP is set at zero energy unless otherwise noted). Deviating from the EP, the energy levels are separated from each other.  We denote the eigenvalues and corresponding eigenvectors of $H(\bm k)$ as $X_n=(z_1,z_2,...,z_n; \psi_1, \psi_2,..., \psi_n)$. As an isolated EP in $d$D can be enclosed by a $(d-1)$D sphere $S^{d-1}$. From the homotopy point of view, classifying the band structures near the EP is to find all the nonequivalent classes of the \textit{non-based} maps from $S^{d-1}$ to $X_n$, denoted as $[S^{d-1},X_n]$. While our main focus here is for isolated EPs, the classification scheme can be directly extended to some other kinds of exceptional degeneracies \cite{Bergholtz1,Bergholtz2,Bergholtz3,Bergholtz4,SM} like exceptional lines in 3D. For generic exceptional band degeneracy of dimension $d_{E}$ ($d_{E}=0,1,2$ corresponds to exceptional point/line/surface, respectively), the topology can be revealed by taking a nearby surface of dimension $D_s=d-d_{E}-1$, with the topological invariants carried by the homotopy invariants $[S^{D_s}, X_n]$.

The eigenvalue part is the configuration space of ordered $n$-tuples $\textrm{Conf}_n(\mathbb{C})$. As each eigenvector is unique up to multiplying a nonzero phase factor, the eigenvector space (denoted as $\Psi$) is $\Psi\cong \textrm{U}(n)/\textrm{U}^n(1)$ \cite{SM}. By further removing the redundancy of permutation of eigenvalues (together with their associated eigenvectors), we have the classifying space $X_n\cong(\textrm{Conf}_n\times\Psi)/S_n$, with $S_n$ the symmetric group of degree $n$. Using the relations $\pi_1(\textrm{Conf}_n)=\textrm{PB}_n$ (pure braiding group) \cite{purebraid1,purebraid2} and $\pi_m(\textrm{Conf}_n)=0$ ($m\geq 2$) \cite{fadell}, and long sequence of homotopy relations \cite{hatcher,class1,class2}, we have:
\begin{eqnarray}
&&\pi_1(X_n)=\textrm{B}_n, ~~~d=2;\label{pi1}\\
&&\pi_2(X_n)=\mathbb{Z}^{n-1},~~~d=3;\label{pi2}\\
&&\pi_{d-1}(X_n)=\pi_{d-1}(\textrm{U}(n)),~~~d\geq 4.\label{pi3}
\end{eqnarray}
For $d=2, n=2$, the homotopy group $\textrm{B}_2\cong\mathbb{Z}$, consistent with the previous characterization of EP$_2$ through vorticity \cite{fuliang} or discriminant number \cite{yangzhesen}. The above homotopy groups are all Abelian except for $d=2, n\geq3$, which is the braid group $\textrm{B}_n$. If we travel along a circle around a 2D EP, the energy levels braid together. Eq. (\ref{pi1}) indicates that the topology of a 2D EP$_n$ is entirely captured by such braiding pattern. The ordinary non-defective point degeneracy has trivial band braidings. Eq. (\ref{pi2}) means a 3D EP$_n$ is characterized by assigning a Chern number to each of the separable $n$ bands, with the constraints that they sum to zero.

The non-based map relates to the homotopy group through the action of the fundamental group \cite{hatcher}:
\begin{eqnarray}
[S^{d-1},X_n]\cong\pi_{d-1}(X_n)/\pi_1(X_n).
\end{eqnarray}
The right-hand side is the orbit set. $[S^{d-1},X_n]$ is not necessary a group and usually decomposed into several distinct sectors, induced by the fundamental group. While the non-based homotopy can be worked out in a case-by-case manner, in 2D, $[S^1,X_n]$ is the conjugacy class of the braid group B$_n$ \cite{class1,class2,knotPRL}. In fact, choosing a different starting point on the encircling path may end up with another braid, which however, conjugate to the original one, hence they are in the same conjugacy class. In 3D, $[S^2,X_n]$ is a collection of $n$ integer Chern numbers. Due to the un-sortability of the complex energy levels, the two integer sets $[s_1,s_2,...,s_n]$ and $[s_{b_1},s_{b_2},...,s_{b_n}]$ (the set induced by the band permutation of any $b\in\pi_1(X_n)$) are identified as the same \cite{class1,class2}. 

{\color{blue}\textit{Knot around EP.}} In the following, we focus on 2D EPs. It turns out there is a one-to-one correspondence between the conjugacy class of braid group $\textrm{B}_n$ and geometric knot in the solid torus \cite{knotbook}, culminating in a \textit{knot} classification of the EPs. Different types of EPs are represented by topologically distinct knots, thus characterized by knot invariants \cite{purebraid2,knotPRL}, e.g., Jones polynomials \cite{Jones}. Intuitively, let us take a closed path $\Gamma(\theta)~(\theta\in[0,2\pi]) $ surrounding the EP. Along $\Gamma$, the $\theta$-dependent energy levels $(z_1,z_2,...,z_n)$ trace $n$ strands which tangle together in the 3D space spanned by (ReE, ImE, $\theta$), as depicted in Fig. \ref{EPknot}(a). The trajectory returns to itself to form a closed knot during the $\theta$-evolution from $0$ to $2\pi$ ($\theta=0$ and $\theta=2\pi$ is identical). The knot topology of a given EP is best represented using braid word. It can be determined by projecting the above energy-level strings onto the $\textrm{Im}E=0$ plane. As $\theta$ evolves, the strings undergo a sequence of crossings. In Artin's notation, we label a crossing as $\tau_i$ ($\tau_i^{-1}$) if the $i$-th string crosses over (under) the $(i+1)$-th string from left. $\tau_i$ satisfy the braid relation: $\tau_i\tau_j=\tau_j\tau_i$ for $|j-i|\geq2$, and $\tau_i\tau_{i+1}\tau_i=\tau_{i+1}\tau_i\tau_{i+1}$. The entire level set is specified by a product of braid operators. For example, $\tau_1^n$ corresponds to twisting two level strands $n$ times; $n=1, 2, 3$ represents unknot, Hopf link, and Trefoil knot, respectively. We note the difference from the 1D knotted separable bands \cite{knotPRL}, here the knots of EPs are attributed to the nontrivial nearby braiding around the band singularities.

The knot near an EP can be regarded as the roots (or nodal set) of the characteristic polynomial (ChP). Denote $z=k_1+ik_2$ and $H(z)=H(k_1,k_2)$, the ChP reads
\begin{eqnarray}\label{chp}
f(\lambda,z)=\det[\lambda-H(z)]=\textstyle{\prod}_{j=1}^n (\lambda-z_j).
\end{eqnarray}
The well-known complex polynomial for the $(p,q)$-torus knot is $f(\lambda,z)=\lambda^p-z^q$, with $p$ roots $z_j=z^{q/p}e^{i2\pi (j-1)/p},~(j=1,2,...,p)$. By tracing a closed path around the EP, the energy levels wind $q$ times around a circle in the interior of the torus, and $p$ times around its axis of rotational symmetry. The simplest Hamiltonian realizing the ($p$,$q$)-torus knot can thus be chosen as:
\begin{eqnarray}\label{torusham}
H_{T_{p,q}}=\left(\begin{array}{cccc}
0 & 0 & 0 & (k_1+i k_2)^q\\
1 & 0 & 0 & 0\\
0 & \ddots & \ddots & 0\\
0 & 0 & 1 & 0
\end{array}\right)_{p\times p}.
\end{eqnarray}
Figs. \ref{EPknot}(b)(c) plot the band structures of EP$_2$ associated with $T_{2,1}$ (unknot) and $T_{2,3}$ (trefoil knot), respectively. The unknot case is the most studied EP$_2$ in the literature. 

To find the Hamiltonian for an EP$_n$ with a prescribed knot (denoted as $\mathcal{K}$), we need to construct the proper ChP $f(\lambda,z)$. As $\lambda,z\in\mathbb{C}$, $f(\lambda,z)$ defines a mapping from $\mathbb{C}^2$ to $\mathbb{C}$. The energy levels $(z_1,z_2,...,z_n)$ are the nodal sets of the ChP and the EP$_n$ is the isolated singularity located at $(\lambda,z)=(0,0)$. Mathematically, the closure of the band braiding near the EP$_n$ is the algebraic knot around the singular point. Akbulut and King \cite{AKing} showed (albeit in a nonconstructive way) that any knot can arise as the knot around such a singular point of a polynomial in the context of complex hypersurfaces. As elaborated in Ref. \cite{SM}, the desired Hamiltonian with the desired knotted EP pattern can be intuitively constructed from the Fourier parameterization of the associated braid diagram. Fig. \ref{EPknot}(d) depicts an EP$_3$ with its nearby energy levels forming a figure-8 knot, described by braid word $(\tau_1^{-1}\tau_2)^2$. Fig. \ref{EPknot}(e) depicts an EP$_4$ associated with $L6a1$ link of braid word $(\tau_1^{-1}\tau_3^{-1}\tau_2)^2$. By further replacing $k_{1,2}$ with $\sin k_{x,y}$ in $H_{\mathcal{K}}(z)$, we get a 2D lattice Hamiltonian hosting the desired EP$_n$ with $\mathcal{K}$ knot topology.

{\color{blue}\textit{Bulk Fermi arc \& discriminant number.}} A direct physical consequence of the EP topology is the appearance of bulk Fermi arcs \cite{fermiarc}. For example, a pair of EPs split from a Dirac point is connected by an open-ended Fermi arc, as identified as the isofrequency contour in photonic experiments \cite{fermiarc}. Here we extend this notion to generic higher-order EPs and define the bulk Fermi arcs as the loci in the ($k_1$, $k_2$) space when any two levels have the same real energy, i.e., Re[$z_i$]=Re[$z_j$] for $i\neq j$. As the EP satisfies this condition, a bulk Fermi arc must start from an EP and end at another; The EPs are the endpoints of Fermi arcs. The constraint imposed by the condition Re[$z_i$]=Re[$z_j$] yields some 1D loci in the $(k_1, k_2)$ space. The bulk Fermi arcs trace open-ended curves emanating from the EPs. From the braid representation, the bulk Fermi arc is formed whenever there is a crossing, either over or under, in the braid diagram. Hence we have:
\begin{eqnarray}\label{crossing}
n_{\textrm{arc}}=n_++n_-=c(\mathcal{K}).
\end{eqnarray}
Here $n_{\pm}$ denotes the number of over/under crossings, respectively. $c(\mathcal{K})$ is the knot crossing number. The number of Fermi arcs $n_{\textrm{arc}}$ emanating from each EP equals to $c(\mathcal{K})$ associated with the EP.

Moreover, the braid crossing dictates the discriminant invariant $\nu$ \cite{yangzhesen}. In its neat form, $\nu$ is the sum of band vorticity \cite{fuliang} for any pair of band:
\begin{eqnarray}
\nu=-\frac{1}{2\pi}\sum_{i\neq j}\oint_{\Gamma}\nabla_{\bm k}\arg(z_i-z_j)\cdot d\bm k.
\end{eqnarray}
Along the closed path $\Gamma$, the EP knot contributes to the vorticity through braid crossings. An over/under crossing contributes a $\pm1$, yielding
\begin{eqnarray}\label{nu}
\nu=n_+-n_-=W(\mathcal{K}),
\end{eqnarray}
where $W(\mathcal{K})\equiv n_+-n_-$ is the writhe of the knot $\mathcal{K}$. It is obvious two different EPs of different braids may have the same discriminant number, e.g., figure-8 and Borromean ring ($L6a4$). We stress that, the full information of an EP is encoded in its knot structure; either the bulk Fermi arc or discriminant number is determined by the braiding and has limited discriminant power to specify an EP. 

\begin{figure}[t]
\centering
\includegraphics[width=3.34in]{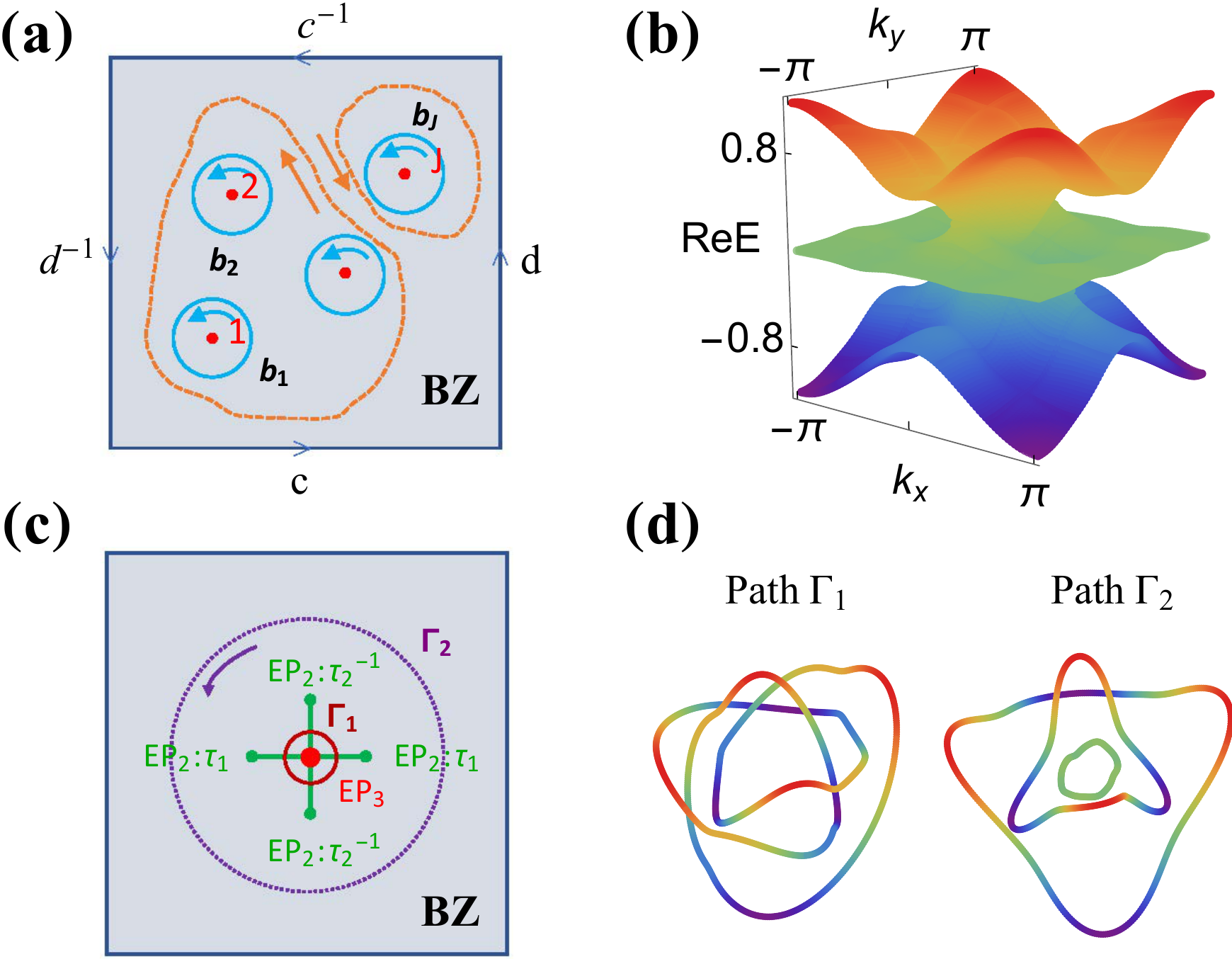}
\caption{Non-Hermitian no-go theorem on a 2D lattice. (a) Sketch of the proof of the theorem. The $j$-th EP ($j=1,2,...,J$) is labeled by level braiding $b_j$ along the nearby path (cyan). The paths can be continuously deformed to the orange curves and finally to the BZ boundary. (b) Band structure of model (\ref{hnogo}). $m=0.3$. (c) EPs and their associated Fermi arcs. The EP$_3$ of figure-8 knot (red dot) at the origin connects with the four neighboring 2-fold EPs (green dots) through bulk Fermi arcs (green lines). (d) Knots formed along the red path $\Gamma_1$ (left panel) and purple path $\Gamma_2$ (right panel) in (c).}
\label{nogo}
\end{figure}

{\color{blue}\textit{No-go theorem.}} The knot classification assigns a specific braiding pattern to a given EP. On a 2D lattice, there may exist multiple different EPs featuring distinct braidings. So what kind of EP configurations is allowed? It turns out the possible EPs on a 2D lattice are governed by the following \textit{no-go theorem}:
\begin{eqnarray}\label{nogotheorem}
b_1b_2...b_J\in[\textrm{B}_n,\textrm{B}_n],
\end{eqnarray}
where $J$ is the total number of isolated EPs in the 2D BZ [see Fig. \ref{nogo}(a)], $b_j$ is the braid for the $j$-th EP along a small surrounding path. $[\textrm{B}_n,\textrm{B}_n]=\{uvu^{-1}v^{-1}|(u, v)\in\textrm{B}_n\}$ is the commutator subgroup. It is easy to check taking a conjugate element of any $b_i$ or switching its order in Eq. (\ref{nogotheorem}) is irrelevant \cite{footnote}. We can continuously expand and deform the enclosing path to the Brillouin zone (BZ) boundary as sketched in Fig. \ref{nogo}(a), without passing through any singularity. The resulting overall band braiding $b_1b_2...b_J$ equals to the band braiding $cdc^{-1}d^{-1}$ along the four edges of the BZ as the upper (left) and lower (right) edge are identical. Specifically, if the two neighboring braidings $c, d$ (or two braidings along any orthogonal line cut of the BZ) commute, the composite braiding must be trivial $b_1b_2...b_J=\textrm{I}$.

The no-go theorem indicates that the braiding of the EP inside the BZ is determined by that of the boundary. It dictates the possible braiding configurations of the isolated EPs and imposes strong constraints on their splittings: the no-go theorem holds for all the isolated split EPs, e.g., a Dirac point splits into two EPs of opposite charges. Note that the commutator subgroup is the set of braids with total exponent sum zero in the braid generator $\tau_i$. According to Eq. (\ref{nu}), the doubling theorem \cite{yangzhesen} naturally arises: the sum of the discriminant number of all EPs vanishes. Hence a single EP of non-zero discriminant invariant must be paired with another one of opposite discriminant number.

We illustrate the no-go theorem through an explicit model:
\begin{eqnarray}\label{hnogo}
H=H_{F8}+m\left(\begin{array}{cc}
c_{x,y} & 0\\
1-c_{x,y}  & -c_{x,y} 
\end{array}\right)\oplus 0_{1\times1},
\end{eqnarray}
with $c_{x,y}=2-\cos k_x-\cos k_y$. Here $H_{F8}$ is the $3\times 3$ lattice Hamiltonian \cite{SM} of the figure-8 EP. Without the second term, $H_{F8}$ hosts four EP$_3$ located at $(0,0)$, $(0,\pi)$, $(\pi,0)$, and $(\pi,\pi)$, respectively. For a finite $m$, only the EP$_3$ at $(0,0)$ survives. A typical band structure is depicted in Figs. \ref{nogo}(b)(c) for $m=0.3$. Besides the EP$_3$ of figure-8 knot with braid word $(\tau_1^{-1}\tau_2)^2$ at the origin, four isolated 2-fold EPs, with braid word $\tau_1$, $\tau_2^{-1}$, $\tau_1$, and $\tau_2^{-1}$, respectively, emerge nearby. Each EP$_2$ connects to the central EP$_3$ through a bulk Fermi arc. The number of Fermi arc emanating from each EP coincides with the knot crossing number [see Eq. (\ref{crossing})]. The knot along a closed path is shown in Fig. \ref{nogo}(d). If the loop is small, enclosing only the central EP$_3$, the energy levels close to a figure-8 knot. In contrast, if the loop encloses all the EPs, the knot is composed of three unlinked components (i.e., trivial braiding) since $\tau_2^{-1}\tau_1(\tau_1^{-1}\tau_2)^2\tau_2^{-1}\tau_1=\textrm{I}$, consistent with the no-go theorem.

{\color{blue}\textit{Discussions.}} To conclude, we have established a homotopy classification of isolated EPs and demonstrated that the knots tied up by energy levels fully characterize the 2D EPs. Based on this new classification scheme, we have demonstrated how to construct NH Hamiltonians corresponding to a given knot. We further proposed a no-go theorem for the EPs on a 2D lattice. Our scheme elegantly relates the bulk Fermi arc and discriminant number to the crossing number and writhe of the knot, respectively. It is worth to mention that the knot topology we presented refers to the knotted structures intrinsic to non-Hermitian EPs from the homotopy perspective, which should not be confused with the usual topological phase in the presence of a well-defined chemical potential. Transitions between different kinds of knotted EPs must be through band touching and rearranging.

The various EP knots and their associated NH Hamiltonians could in principle be realized in platforms such as photonic lattice \cite{nhwsmhhp,longhi,cavity1} or electric circuits \cite{RLCmeasure1,RLCmeasure2,zhao,RLC3,knotmomentum}. For the former, the asymmetric coupling between the ring resonators can be implemented via auxiliary microring cavities. The consequent bulk Fermi arcs should be extracted through resonances in frequencies \cite{fermiarc}. For the latter, the NH Hamiltonian is simulated by the circuit Laplacian, with its band structures given by the admittance spectra. We note that the model Hamiltonian in Eq. (\ref{hnogo}) no-go theorem is merely chosen for illustrative purpose of the no-go theorem and the figure-8 EP therein requires the fine tuning of parameters. However, the stability of such knotted EPs can in fact, be ensured by e.g., PT, CP, pseudo-Hermiticity, pseudo-chirality symmetries \cite{epdelplace,epmandal} due to the symmetry reduction of constraints of EPs. Under a symmetry-breaking perturbation, the knotted EP may locally split into other types of band singularities; however, the geometry knot formed by encircling the overall band singularities after splitting remains unchanged, as dictated by the no-go theorem. The various knotted EPs discovered here should naturally emerge as the critical point of the nonreciprocal phase transition \cite{nrpt}. Beyond the isolated EPs considered here, more intricate line- or surface-degeneracies may exist. For example, in 2D, encircling the EP line yields the so-called singular knots. It would be interesting to extend the analysis to such singular knots and reveal their physical consequences.
\begin{acknowledgments}
This work is supported by the NSFC under Grants No.11974413 and the Strategic Priority Research Program of Chinese Academy of Sciences under Grant No. XDB33000000.
\end{acknowledgments}

\clearpage
\onecolumngrid
\appendix

\section{Supplementary Materials}

In this Supplementary Material, we provide details on (\textrm{I}) the classification scheme of exceptional degeneracy from homotopy perspective; (\textrm{II}) the homotopy equivalence between space $U(n)/U^n(1)$ and $GL(n,\mathbb{C})/C^{*n}$; and (\textrm{III}) the construction of the characteristic polynomial and its associated Hamiltonian corresponding to a prescribed knotted exceptional point (EP).

\subsection{(I) Classification scheme of exceptional degeneracy from homotopy perspective}
The classification scheme for isolated EPs can be extended to more generic exceptional degeneracies like exceptional lines and surfaces. Intuitively, the various types of exceptional degeneracies can be regarded as topological ``defects" in the parameter space. The exceptional point/line/surface is taken as point/line/surface defect, respectively. In a similar vein as the characterization of defects in topological insulators \cite{defect1}, the strategy here is to take a small nearby surface enclosing the considered exceptional degeneracy and to seek the topological invariants on the enclosing surface. Let us suppose a $d_{E}$-dimensional exceptional degeneracy ($d_{E}=0,1,2$ corresponds to exceptional point/line/surface) in $d$ dimensional parameter space. The enclosing surface is then chosen as $D_s=d-d_{E}-1$, and the exceptional degeneracy is governed by the homotopy invariant
\begin{eqnarray}
[S^{D_s},X_n],
\end{eqnarray}
i.e., the nonequivalent class of the non-based maps from $S^{D_s}$ to $X_n$. Here $S^{D_s}$ is the $D_s$-sphere, $X_n$ is the space of $n$-band non-Hermitian Hamiltonians as defined in the main text. It is clear that an isolated EP ($d_{E}=0$) is characterized by the homotopy invariant $[S^{d-1},X_n]$.

\begin{figure}[b!]
\centering
\includegraphics[width=0.75\textwidth]{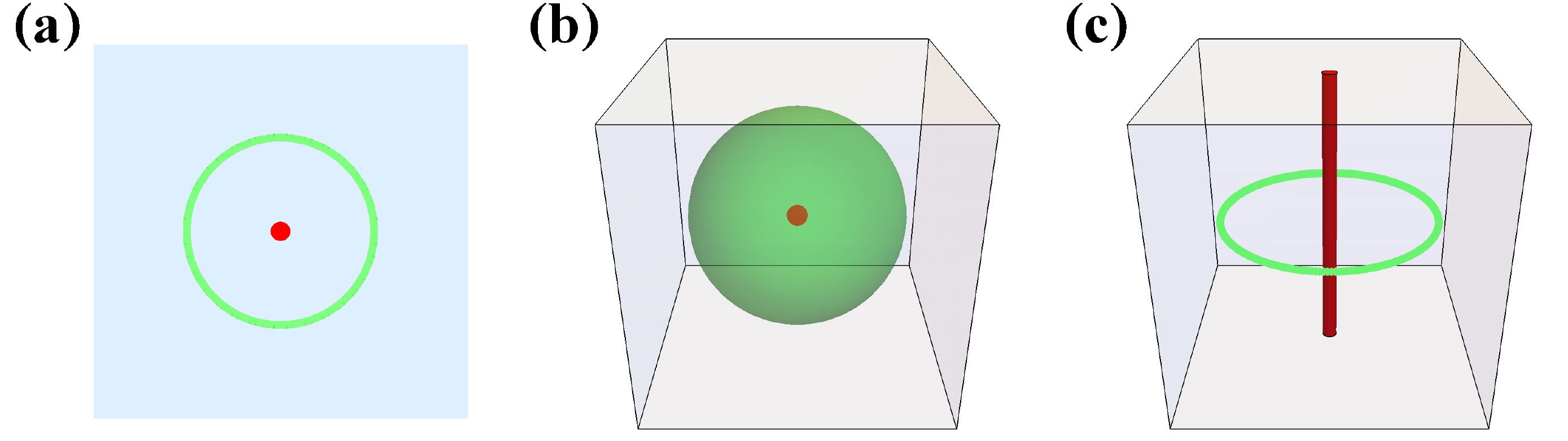}
\caption{Sketch of some typical exceptional degeneracies and their homotopy characterizations. (a) Exceptional point in 2D, $(d,d_E)=(2,0)$. (b) Exceptional point in 3D, $(d,d_E)=(3,0)$. (c) Exceptional line in 3D, $(d,d_E)=(3,1)$. The exceptional degeneracies are marked in red and the enclosing surface is marked in green.}\label{fs1}
\end{figure}

Several typical exceptional degeneracies in 2D and 3D are sketched in Fig. \ref{fs1}. Fig. \ref{fs1}(a) is for an isolated EP in 2D. It is characterized by the homotopy invariant $[S^1,X_n]$. In the main text, $[S^1,X_n]$ is identified as the conjugacy class of braid group $B_n$ and further one to one corresponds to the geometric knot formed by the eigenenergies nearby the EP. Fig. \ref{fs1}(c) is for an exceptional line (red line) in 3D. It topology is carried by the encircling loop (green) around the. In this case $d=3, d_E=1$ and $D_s=1$. It is also characterized $[S^1,X_n]$, or conjugacy class of braid group. The same conclusion can be obtained for an exceptional surface in 4D parameter space with $d=4, d_E=2$. Fig. \ref{fs1}(b) is for an EP in 3D with $d=3, d_E=0$ (similarly for an exceptional line in 4D). It is characterized by $[S^2,X_n]$, which is a collection of Chern numbers as discussed in the main text. In the presence of certain symmetries, more intriguing defects like, exceptional line in 2D ($d_{E}=1,d=2$) and exceptional surface ($d_{E}=2,d=3$) in 3D may appear. For these cases, $D_s=0$. The zero-th homotopy group $\pi_0 (X_n )$, the set of path components of the space $X_n$, does not reveal too much information. The stability of such band singularities should be analyzed in a case-by-case manner, e.g., using the resultant vector as in \cite{epdelplace}.

\subsection{(II) The homotopy equivalence between space $U(n)/U^n(1)$ and $GL(n,\mathbb{C})/\mathbb{C}^{*n}$}
For non-Hermitian Hamiltonian with $n$ separable bands (i.e., $E_m\neq E_n$ for any $m\neq n$ with some fixed parameters), the space of the eigenvectors is $GL(n,\mathbb{C})/\mathbb{C}^{*}$ \cite{class2}. Here $GL(n,\mathbb{C})$ comes from $n$ linearly independent complex vectors, $\mathbb{C}^*=\mathbb{C}-\{0\}$ comes from the redundancy of biorthogonal rescaling by some nonzero complex numbers \cite{coll5}. In the following, we showcase that the eigenvector space $GL(n,\mathbb{C})/\mathbb{C}^{*n}$ is homotopy equivalent to $U(n)/U^n (1)$ in the main text, with $U(n)$ the $n\times n$ unitary matrices. Compared to the general linear group, the unitary group is more commonly used in the context of algebraic topology. The equivalence is due to the two facts: (1) $U(n)$ is a deformation retraction \cite{hatcher} of $GL(n,\mathbb{C})$; (2) The punctured complex plane $\mathbb{C}^*\cong S^1\cong U(1)$, ($\cong$ means homeomorphic, it preserves all the topological properties of two spaces). While the first statement can be explicitly visualized through the Gram-Schmidt process, which takes $n$ linearly independent vectors into $n$ orthogonal vectors in a continuous way, here we give an exact proof: up to homeomorphism, $GL(n,\mathbb{C})\cong U(n)\times \mathbb{C}^{\frac{n(n+1)}{2}}$. 

That being said is a direct consequence of the polar decomposition and Cholesky decomposition in linear algebra.\\

\textit{Polar decomposition}

	For any real (complex) $n\times n$ invertible matrix $A$, there is a unique orthogonal (unitary) matrix $R$ and positive-definite symmetric matrix $S$ so that $A=RS$.	\\

\textit{Cholesky decomposition}

	Any symmetric positive definite matrix $S$ can be written uniquely as $S=LL^{\dagger}$ where $L$  is a lower-triangular matrix with strictly positive diagonal entries.

Further note that there are $\frac{n(n+1)}{2}$ nonzero elements in an $n\times n$ lower-triangular matrix and $(0,\infty)\cong(-\infty,\infty)$ up to homeomorphism. Hence statement (1) is proved. Since $\mathbb{C}$ is contractible to a point (with trivial homotopy groups), it can be neglected from the homotopy perspective. It follows that the homotopy groups of the above two spaces $U(n)/U^n(1)$ and $GL(n,\mathbb{C})/\mathbb{C}^{*n}$ are the same. 

\subsection{(III) Construction of the characteristic polynomial and Hamiltonian associated with a given knotted exceptional point}
In the  main text, we have demonstrated that any isolated EP$_n$ can be assigned a knot structure $\mathcal{K}$ due to the exact correspondence between the geometric knot and the conjugacy class of braid group. Here we showcase how to construct the characteristic polynomial and Hamiltonian associated with an EP$_n$ with desired braiding pattern (or knot structure $\mathcal{K}$). Mathematically, the closure of the band braiding near the EP$_n$ is the algebraic knot around the singular point. 

The characteristic polynomial $f(\lambda,z)=\det(\lambda-H(k_1,k_2))$ with $\lambda,z=k_1+i k_2\in\mathbb{C}$, defines a mapping from $\mathbb{C}^2$ to $\mathbb{C}$. The energy levels $(z_1,z_2,...,z_n)$ are the nodal sets of $f(\lambda,z)=0$ and the EP$_n$ is the isolated singularity located at $(\lambda,z)=(0,0)$. $f(\lambda,z)=0$ is semiholomorphic and can be written as a complex polynomial in $\lambda$, $z$, and $\bar{z}$. To proceed, we need an auxiliary polynomial $f_{\theta}:\mathbb{C}\times S^1\rightarrow\mathbb{C}$, with the knot $\mathcal{K}$ as its nodal set. This process has been elaborated in \cite{knotPRL,bode}. The key lies in the Fourier parameterization of the braid diagram associated with the knot. Having obtained $f_{\theta}$, we can replace $e^{\pm i\theta}$ with $z$ and $\bar{z}$, respectively: $f(\lambda,z)=f_{\theta}(e^{i\theta}\rightarrow z;e^{-i\theta}\rightarrow\bar{z})$. In some cases, however, such an analytical continuation does not necessarily guarantee the EP$_n$ is the isolated singular point of $f(\lambda,z)$. To overcome this issue, we slightly modify the characteristic polynomial to take the following form ($g>0$ is some integer) \cite{bode2}:
\begin{eqnarray}\label{fmodify}
f(\lambda,z)=(z\bar{z})^{ng} f(\frac{\lambda}{(z\bar{z})^g},\frac{z}{\sqrt{z\bar{z}}}).
\end{eqnarray}

Through the above procedures, we get the characteristic polynomial of e.g., the figure-8 knot: $f_{F8}=1/64[64\lambda^3-12\lambda a^2[3(z\bar{z})^2+2z^2(z\bar{z})-2\bar{z}^2(z\bar{z})]-14a^3(z^2+\bar{z}^2)(z\bar{z})^2-a^3(z^4-\bar{z}^4)(z\bar{z})]$ with $a=1/4$. Similarly , we can obtain the characteristic polynomial for more complicated knots, e.g., the $L6a1$ link of braid word $(\tau_1^{-1}\tau_3^{-1}\tau_2)^2$.

To obtain the associated non-Hermitian Hamiltonian, we expand the characteristic polynomial in the series of $\lambda$, which takes the form: $f(\lambda,z)=\lambda^n-\sum_{j=0}^{n-1}\zeta_j(z) \lambda^j$. It is clear at the location of EP$_n$, $z=0$, $\zeta_j(z)=0$. A natural choice of the knotted Hamiltonian is
\begin{eqnarray}\label{hk}
H_{\mathcal{K}}(z)=\left(\begin{array}{cccc}
\zeta_{n-1}(z)  &~~~ ...~~~& \zeta_{1}(z) & \zeta_{0}(z)\\
1  & 0 & 0 & 0\\
0 & \ddots & \ddots & 0\\
0  & 0& 1 & 0\\
\end{array}\right)_{n\times n}.
\end{eqnarray}
Obviously $\det[\lambda-H_{\mathcal{K}}(z)]=f(\lambda,z)$. Fig. 1 of the main text showcases several typical non-Hermitian bands with the knotted EPs located at the origin through the above procedures. On a 2D lattice, we simply replace $k_{1,2}$ with $\sin k_{x,y}$ in $H_{\mathcal{K}}(z)$ and get a 2D lattice Hamiltonian hosting the desired EP$_n$ of the $\mathcal{K}$ knot. 


\begin{thebibliography}{99}
\bibitem{footnote} This is due to the fact that for any $x, y\in\textrm{B}_n$, (1) $x[u,v]x^{-1}=[xux^{-1},xvx^{-1}]\in[\textrm{B}_n,\textrm{B}_n]$; (2) $yb_1y^{-1}b_2...b_J=(yb_1y^{-1}b_1^{-1})(b_1b_2...b_J)=[y,b_1][u,v]\in[\textrm{B}_n,\textrm{B}_n]$; and (3) $b_{i+1}b_i=b_i^{-1}(b_ib_{i+1})b_i$.
\end{thebibliography}
\end{document}